\documentclass[prb,twocolumn,superscriptaddress]{revtex4}
\usepackage{dcolumn}
\usepackage{bm}% bold math

\begin{document}

\title{Dopants adsorbed as single atoms prevent degradation of catalysts}
       
\author{Sanwu~Wang}
\affiliation{Department of Physics and Astronomy,
       Vanderbilt University, Nashville, Tennessee 37235, USA.}
\author{Albina~Y.~Borisevich}
\affiliation{Condensed Matter Sciences Division, Oak Ridge National
       Laboratory, Oak Ridge, Tennessee 37831, USA}      
\author{Sergey~N.~Rashkeev}
\affiliation{Department of Physics and Astronomy,
       Vanderbilt University, Nashville, Tennessee 37235, USA.}
\author{Michael~V.~Glazoff}
\affiliation{Alcoa Technical Center, Alcoa Center, Pennsylvania
       15069-0001, USA}      
\author{Karl~ Sohlberg}
\affiliation{Department of Chemistry, Drexel University, Philadelphia,
       Pennsylvania 19104, USA}
\author{Stephen~J.~Pennycook}
\affiliation{Condensed Matter Sciences Division, Oak Ridge National
       Laboratory, Oak Ridge, Tennessee 37831, USA}
\author{Sokrates~T.~Pantelides}
\affiliation{Department of Physics and Astronomy,
       Vanderbilt University, Nashville, Tennessee 37235, USA.}
\affiliation{Condensed Matter Sciences Division, Oak Ridge National
       Laboratory, Oak Ridge, Tennessee 37831, USA}

\begin{abstract}

\end{abstract}

\maketitle

%doi:10.1038/nmat1077

{\bf The design of catalysts with desired chemical and thermal properties
is viewed as a grand challenge for scientists and engineers\cite{1}. For 
operation at high temperatures, stability against structural
transformations is a key requirement. Although doping has been found to 
impede degradation, the lack of atomistic understanding of the pertinent
mechanism has hindered optimization. For example, porous
{$\gamma$}-Al$_2$O$_3$, a widely used catalyst and catalytic
support\cite{2,3,4,5,6}, transforms to non-porous {$\alpha$}-Al$_2$O$_3$
at $\sim$1,100~$^\circ$C (refs~\onlinecite{7,8,9,10}). Doping with La 
raises the transformation temperature\cite{8,9,10,11} to
$\sim$1,250~$^\circ$C, but it has not been possible to establish if La
atoms enter the bulk, adsorb on surfaces as single atoms or clusters, or
form surface compounds\cite{10,11,12,13,14,15}. Here, we use direct
imaging by aberration-corrected Z-contrast scanning transmission
electron microscopy coupled with extended X-ray absorption fine structure
and first-principles calculations to demonstrate that, contrary to
expectations, stabilization is achieved by isolated La atoms adsorbed on
the surface. Strong binding and mutual repulsion of La atoms effectively
pin the surface and inhibit both sintering and the transformation to
{$\alpha$}-Al$_2$O$_3$. The results provide the first guidelines for the
choice of dopants to prevent thermal degradation of catalysts and other
porous materials.}

Aberration correction has allowed sub-{\AA}ngstrom beams to be formed with
huge improvements in the visibility of single atoms\cite{16}. Figure 1a   
shows a Z-contrast image of a flake of La-doped {$\gamma$}-Al$_2$O$_3$ in   
the [100] orientation. Surprisingly, unlike the undoped case\cite{17,18},
there was no apparent preferential exposure of the [110] surface. Instead,
the [100] surface was encountered often. The square arrangement of Al--O
columns is clearly resolved. In this imaging mode, the intensity
contributed by an atom is roughly proportional to Z$^2$, where Z is the
atomic number\cite{19}, allowing single La atoms to be visible in the form
of brighter spots on the background of thicker but considerably lighter 
{$\gamma$}-Al$_2$O$_3$ support. For the average sample thickness, the
nominal density (3.5~wt\%) corresponds to $\sim$3 La atoms per nm$^2$ in
projection. Typical counts of La atoms observed in scanning transmission
electron microscopy (STEM) images are consistent with that density. Most
of the La atoms are located directly over Al--O columns (site A in Fig.
1a), but a small fraction also occupies a position shifted from the Al--O
column (site B in Fig. 1a). Note that in this second configuration, the La
atom appears much dimmer, because its intensity is no longer superimposed
on that of the Al--O column; this effect is also evident in simulated
images using multislice codes\cite{20} (Fig. 1b) and in the line traces
of the experimental data (Fig. 1c). Furthermore, substitutional La atoms
on the beam entrance side of the alumina flake are systematically less 
bright than those on the exit face (site A$'$ on Figs 1a--c), where
channelling has sharpened the beam; for interstitial La atoms (site B) the
intensity difference in the simulated image (Fig. 1b) between the entrance
(arrow up) and exit (arrow down) surface is negligible. The images reveal
clearly that there is no correlation in the distribution of dopant atoms.
The presence of La was also confirmed using electron energy-loss spectra
(EELS), acquired over 100-nm$^2$ areas of the sample. Atomically resolved
EELS was not feasible for this material because the necessary prolonged
exposure caused significant damage.

Although a Z-contrast image is a two-dimensional (2D) projection of the 3D
object, the higher convergence angle of the aberration-corrected STEM
probe offers a possibility of depth sensitivity using the decreased depth
of focus. Under these conditions, the intensity of an image of a point
object (such as single atom) significantly diminishes when it is only
nanometres away from the perfect focus. We find that for La-doped
{$\gamma$}-Al$_2$O$_3$ flakes in the absence of any lattice contrast from
the substrate (that is, when it is tilted far off crystallographic axes)
two different focus conditions can be identified, which can a gain be
attributed to La atoms located on the top and bottom surface of the flake,
respectively. Figure 2a shows an example of such through-focal series of
Z-contrast STEM images. It is clear from the figure that the bright spots
corresponding to La atoms are sharper at defocus values of 0 and $-$8 nm,
suggesting that the thickness of the examined flake is close to 8 nm. Note
that after going past the bottom surface, atoms `fade' faster, reflecting
widening of the probe after passing through the substrate. This conclusion
can also be expressed quantitatively as the dependence of median bright 
spot intensity on defocus (Fig. 2b). This unique experiment demonstrates a
depth-sensing capability of the STEM, and corroborates the conclusion that
La atoms maintain widely separated surface positions even after
high-temperature annealing. Figure 2a also shows that La atoms move
relative to one another as the imaging process is repeated with different
defocus. Such motion is consistent with the fact that the La atoms reside
on the surfaces, where the beam can induce their migration.

Extended X-ray absorption fine structure (EXAFS) measurements were
performed on a series of five lanthanum-doped {$\gamma$}-alumina
specimens: a sample without heat treatment (sample 0) and 4 samples
annealed at 800, 1,000, 1,200, and 1,400~$^\circ$C for 3 hours (samples
1--4, respectively). The Fourier transforms of the X-ray absorption
spectra recorded at the La L$_{\text{III}}$-edge (Fig. 3.) demonstrate
that up to 1,200~$^\circ$C (that is, before the collapse of
{$\gamma$}-Al$_2$O$_3$ structure) no noticeable changes occur in the
coordination of La. Simulations of the EXAFS spectra differ markedly from
the observed spectra if one or more La atoms are inserted in the first or
second coordination shells of La, indicating that adsorbed La atoms are
isolated, in agreement with the conclusion reached from the Z-contrast    
images. An excellent fit is obtained for sample 4 by using twelve oxygen
atoms in the first coordination shell and six aluminium atoms in the
second shell, suggesting an ordered bulk phase. Simulation of the spectra
from samples 0--3 show that the peak at $\sim$3~{\AA} of sample 4 can be
suppressed by decreasing the number of oxygen atoms in the first shell
from twelve to less than eight, and distributing La--O distances,
suggesting surface-like environment. The significant change in the local
coordination of the La occurring at 1,400~$^\circ$C (sample 4) can be
attributed to the {$\gamma$}--{$\alpha$} phase transition. The exact form
of La after the transition has not yet been identified.

Figure 4 represents the minimum energy configurations of the undoped and  
La-doped (100) surface obtained from first-principles calculations. In the
absence of La, the (100) surface (Fig. 4a) shows only minor relaxation
effects; all the surface aluminium atoms are five-coordinated and the
surface oxygen atoms are either three- or four-coordinated. The cation
vacancies are located between the first and second oxygen subsurface
layer. However, when a La atom is introduced, a significant relaxation of
the structure occurs. One of the five-coordinated Al atoms (adjacent to      
La) is displaced from the surface into the subsurface tetrahedral vacancy
site (Fig. 4b). The La atom occupies the resultant surface four fold
hollow site, which is close to the initial location of the Al atom in
planar coordinates but located $\sim$1.2~{\AA} above it, making La--O bond
lengths 2.3--2.5~{\AA}. This configuration, obtained independently by
total-energy minimization, is precisely the same as site A on the
micrograph in Fig. 1a. Calculations aimed at revealing the structure of
the observed B site found that it corresponds to a La atom with one of the
four neighbouring surface oxygen atoms missing. The asymmetry forces the
La atom off the Al--O column, as observed. Its formation is clearly the
result of the presence of surface oxygen vacancies.

Additional calculations exploring the incorporation of La atoms in bulk  
{$\gamma$}-Al$_2$O$_3$ found that a La atom, when initially placed at a   
vacancy, interstitial, or substitutional site in the second or third   
subsurface layer, would relax up to the surface. When a La atom is
initially located in a deeper layer (the 8th or 9th layer of the supercell
for the (100) surface, and the 5th or 6th layer for the (110C) surface),
which is equivalent to the bulk, the total energy of the system is
significantly higher than that of the configuration with La on the surface
(models with La in different bulk sites were examined). The preference for
surface sites over the bulk arises primarily from the strong binding of La
on the surface and the large difference in ionic size between La$^{+3}$   
(1.03~{\AA}) and Al$^{+3}$ (0.54~{\AA})\cite{21}. The theoretical result
corroborates the through-focus imaging analysis of Fig. 2. The marked
preference for surface sites versus the bulk is an important factor in the
inhibition of sintering. Progress of the sintering process would
inevitably trap some of the surface La atoms in the bulk, thus forcing the
system into highly strained and energetically unfavourable configuration.
The resulting inhibition effect helps retain a large specific surface area   
for {$\gamma$}-Al$_2$O$_3$ at higher temperatures, in agreement with
experimental observations\cite{9,10,11,22}.

The binding energy of La to the (100) surface is very high (8.6~eV), due
partly to the removal of the surface Al atom into the subsurface, which
enhances the attractive interaction between La and the surface oxygen  
atoms and reduces the otherwise strong repulsion between La and the Al
atom. The strong binding also causes large migration energies (4--5~eV)
for typical paths connecting equivalent configurations. Similar
calculations for the (110C) surface, which is exposed preferentially in
the undoped {$\gamma$}-Al$_2$O$_3$ (refs~\onlinecite{17,18}), also
resulted in high values of the binding energy (7.5~eV). In this case,
however, La atoms occupy existing surface hollow sites, which are created
on the undoped surface\cite{18,23} by displacement of three-coordinated
surface Al atoms into the empty octahedral sites in the first subsurface
layer. The difference in the two binding energies explains the occurrence
of (100) surfaces after annealing in the presence of La dopant. When the
same computational procedure is carried out for the {$\alpha$}-Al$_2$O$_3$
(0001) surface, much lower binding energy (4.3~eV) for La atoms is
obtained. This difference means that doping would cause an in crease in
the enthalpy of {$\alpha$}-Al$_2$O$_3$ relative to {$\gamma$}-Al$_2$O$_3$,
thus further stabilizing {$\gamma$}-Al$_2$O$_3$.

Stabilization of {$\gamma$}-Al$_2$O$_3$ by La is often suggested to be due
to LaAlO$_3$ or La$_2$O$_3$ monolayers on the surface, and the debate
about the formation of one phase over another is still ongoing%
\cite{10,11,12,13,14,15}. We explored the possibility of clustering of
surface La atoms. When placed in the nearest interstitials above the
(110C) surface of {$\gamma$}-Al$_2$O$_3$, two La atoms do not show any
tendency to create any bond between them. Instead, they move away from
each other when the initial distance between them is less than 4~{\AA},
thus suggesting that there is no driving force for the formation of dopant
clusters or monolayers, unlike previously proposed models%
\cite{12,13,14,15} and in complete agreement with our images. This effect
suggests that the sintering can be effectively inhibited by a very small
amount of La dopant, provided that atomically dispersed distribution can
be achieved by a given preparation method; indeed laboratory experiments
demonstrate that stabilization can be achieved with only 0.3~wt\% of
dopant versus the usual 3--5\% (ref.~\onlinecite{10}).

To summarize, we have used a combination of experimental and theoretical
results to elucidate the role of La impurities in the stabilization of  
{$\gamma$}-Al$_2$O$_3$ to high temperatures.  The three methods yield   
complementary information that leads to a single overriding conclusion: La
atoms eschew the bulk and adsorb strongly on {$\gamma$}-Al$_2$O$_3$
surfaces as isolated atoms, strongly pinning the surface and impeding
sintering and phase transformation to avoid getting trapped in the bulk of
{$\gamma$}-Al$_2$O$_3$ or the surface of {$\alpha$}-Al$_2$O$_3$. These
results can be used as guidelines in the choice of dopants for other  
systems where thermal stability is important.

\section*{METHODS}

Samples were prepared using commercial material, Alcoa product Ga-200L,
which is {$\gamma$}-Al$_2$O$_3$ stabilized with La (3.5~wt\% of  
La$_2$O$_3$). The samples for the STEM study were annealed at
1,000~$^\circ$C for 10 hours, which produced higher crystallinity,
facilitating Z-contrast imaging.

Z-contrast STEM observations were made with VG Microscopes HB603U (East
Grinstead, UK) operated at 300~kV and equipped with a Nion (Kirkland,  
Washington, USA) aberration corrector to give a probe size of
0.7--0.8~{\AA} and superior signal-to-noise ratio\cite{24}.

EXAFS measurements were performed in the regime of total external
reflection at room temperature at the Institute of Nuclear Physics of the
Russian Academy of Sciences (the Novosibirsk Center for Synchrotron
Radiation). Synchrotron radiation was properly pre-conditioned using the
Si (111) monochromator. FEFF5 software was used for EXAFS data analysis.

The first-principles calculations were performed within density functional
theory, using the pseudopotential method and a plane-wave basis
set\cite{25}. Exchange correlation was included using the generalized 
gradient-corrected functionals (GGA) given by Perdew and
Becke\cite{26,27}. We adopted the Vanderbilt ultrasoft pseudopotentials   
for oxygen and hydrogen atoms, a norm-conserving pseudopotential for Al,
and a projector augmented wave (PAW) potential for La
(refs~\onlinecite{28,29,30}). A plane-wave energy cut-off of 400 eV and   
two special {\bf k} points in the irreducible part of the 2D Brillouin
zone of the surfaces were used for calculating both the (100) and (110)
surfaces of {$\gamma$}-Al$_2$O$_3$. Semi-infinite (100) and (110) surfaces
were modelled by repeated slabs (supercells) containing 7--12 atomic
layers for the (100) surface and 5--8 layers for the (110) surface
(68--112 and 70--128 atoms, respectively) separated by a vacuum region
equivalent to 10--12~{\AA} (supercells containing 80 atoms and a 
$2~\times~2$ surface cell were used to represent the
{$\alpha$}-Al$_2$O$_3$ surface). The cation vacancies that are inherently
present in the spinel form of {$\gamma$}-Al$_2$O$_3$ were located on the
tetrahedral cation sublattice. All the atoms in the supercell except for
those in the lower one or two at omic layers (which were kept fixed) were
relaxed until the forces on the atoms were smaller than 0.05~eV/{\AA}.

\begin{figure}[h]
\caption{{\bf La atoms on $\gamma$-Al$_2$O$_3$ flake in [100] orientation.
a,} Z-contrast STEM image. Two distinct sites (A(A$'$) and B) for individual
La atoms are seen (images are high- and low-pass filtered). {\bf b,}
Multislice simulation\cite{20} of A, A$'$ and B positions of La atoms over
and under a 40--{\AA} slab of {$\gamma$}-Al$_2$O$_3$ at zero defocus for
an aberration-free probe of 15~mrad semi-angle; one phonon configuration
was found sufficient to achieve convergence. {\bf c,} Intensity (in
arbitrary units, a.u.) profiles of the smoothed raw data corresponding to
the line traces denoted on {\bf a.}}
\end{figure}

\begin{figure}[h]
\caption{{\bf Depth-sensitive STEM experiment. a,} STEM images
(identically processed) taken at different defocus values. Note the
intensity peaking at defocus values of 0 and $-$8~nm. Focus drift did not
exceed 0.4~nm from the first to the last frame in the series. Bar length
is 1~nm. {\bf b,} Median incremental intensity of the La atom images
versus defocus (see Supplementary Information for the original statistical
data).}
\end{figure}

\begin{figure}[h]
\caption{{\bf Fourier transforms (into direct space) of the synchrotron
EXAFS La L$_{\text{III}}$-edge spectra $\chi(k)$ (where $k$ is the wave
vector) of La/{$\gamma$}-Al$_2$O$_3$ samples ($R$ denotes distance from La
atom).} Sample 0 without heat treatment, and samples 1--4 annealed at 800,
1,000, 1,200, and 1,400~$^\circ$C, respectively. Note that no changes in
the spectra occur before 1,400~$^\circ$C. The peaks below 1.5~{\AA} are
artefacts arising from a Fourier transform on a finite domain.}
\end{figure}

\begin{figure}[h!]
\caption{{\bf Configurations for the (100) surface of
{$\gamma$}-Al$_2$O$_3$, determined by first-principles calculations. a,}
undoped, and {\bf b,} La-doped. The Al, O, H and La atoms are shown in
grey, red, white, and blue, respectively. The position of the aluminium
atom that relaxes from the surface into the cation vacancy is indicated by
the blue arrow.}
\end{figure}

\acknowledgments

We thank Valeria V. Vavilova and Yanzhao Cao for their help and John W.
Novak, Jr for his support. Access to the Novosibirsk Synchrotron Facility
(Russian Academy of Sciences) and to the Florida State University
supercomputers (csit1 and csit2) are also gratefully acknowledged. This
work was supported in part by National Science Foundation Grant
DMR-0111841, Department of Energy Grant at Drexel University
DE-FC02-01CH11085, by the William A. and Nancy F. McMinn Endowment at
Vanderbilt University, by the Division of Materials Sciences, by
Laboratory Directed R\&D Program for the US Department of Energy, under
contract DE-AC05-00OR22725 managed by UT-Battelle, and by an appointment
to ORNL Postdoctoral Research Program administered jointly by ORNL and Oak
Ridge Institute for Science and Education (ORISE).


\begin{references}

\bibitem{1} 
Bell, A.T. The impact of nanoscience on heterogeneous catalysis. {\it
Science} {\bf 299,} 1688-1691 (2003).

\bibitem{2}
Satterfield, C.N. {\it Heterogeneous Catalysis in Practice} (McGraw Hill,
New York, 1980).

\bibitem{3}
Knozinger, H. \& Ratnasamy, P. Catalytic aluminas  surface models and
characterization of the surface sites. {\it Catal. Rev. Sci. Eng.} {\bf
17,} 31--69 (1978).

\bibitem{4}
Gates, B.C. Supported metal clusters: synthesis, structure, and catalysis.
{\it Chem. Rev.} {\bf 95,} 511--522 (1995).

\bibitem{5}
Xu, Z. et al. Size-dependent catalytic activity of supported metal
clusters. {\it Nature} {\bf 372,} 346--348 (1994).

\bibitem{6}
Lin, V.S.Y. et al. Oxidative polymerization of 1,4-diethynylbenzene into
highly conjugated poly(phenylene butadiynylene) within the channels of  
surface-function alized mesoporous silica and alumina materials. {\it J.
Am. Chem. Soc.} {\bf 124,} 9040--9041 (2002).

\bibitem{7}
Wefers, K. \& Misra, C. Oxides and hydroxides of aluminum. (Alcoa
Technical Paper No.19, Alcoa Laboratories, Pittsburgh, 1987).

\bibitem{8}
Arai, H. \& Machida, M. Thermal stabilization of catalyst supports and
their application to high-temperature catalytic combustion. {\it Appl.
Catal. A} {\bf 138,} 161--176 (1996).

\bibitem{9}
Church, J.S., Cant, N.W. \& Trimm, D.L. Surface area stability and
characterization of a novel sulfate-based alumina modified by rare earth
and alkaline earth ions. {\it Appl. Catal. A} {\bf 107,} 267--276 (1994).

\bibitem{10}
Glazov, M.V., Novak, J. \& Hector, L.G. Stabilization of
{$\gamma$}-alumina: how much lanthanum do we actually need? (Alcoa
Technical Paper No. 99--168, Alcoa Technical Center, Pittsburgh, 1999).

\bibitem{11}
Kumar, K.-N.P., Tranto, J., Kumar, J. \& Engell, J.E. Pore-structure
stability and phase transformation in pure and M-doped (M = La, Ce, Nd,
Gd, Cu, Fe) alumina membranes and catalyst supports. {\it J. Mater. Sci.
Lett.} {\bf 15,} 266--270 (1996).

\bibitem{12}
Oudet, F., Courtine, P. \& Vieux, A. Thermal stabilization of transition
alumina by structural coherence with lanthanide aluminum oxide (LnAlO$_3$,
Ln = lanthanum, praseodymium, neodymium). {\it J. Catal.} {\bf 114,}
112--120 (1988).

\bibitem{13}
Beguin, B., Garbowski, E. \& Primet, M. Stabilization of alumina by
addition of lanthanum. {\it Appl. Catal.} {\bf 75,} 119--132 (1991)

\bibitem{14}
Vazquez, A. et al. X-ray diffraction, FTIR, and NMR characterization of
sol--gel alumina doped with lanthanum and cerium. {\it J. Solid State
Chem.} {\bf 128,} 161--168 (1997).

\bibitem{15}
Yamamoto, T., Tanaka, T., Matsuyama, T., Funabiki, T. \& Yoshida, S.
Structural analysis of La/Al$_2$O$_3$ catalysts by La K-edge XAFS. {\it J.
Synchrotron Rad.} {\bf 8,} 634--636 (2001).

\bibitem{16}
Batson, P.E., Dellby, N. \& Krivanek, O.L. Sub-{\aa}ngstrom resolution   
using aberration corrected electron optics. {\it Nature} {\bf 418,}
617--620 (2002).

\bibitem{17}
Blonski, S. \& Garofalini, S.H. Molecular dynamics simulations of
{$\gamma$}-alumina and {$\alpha$}-alumina surfaces. {\it Surf. Sci.} {\bf
295,} 263--274 (1993).

\bibitem{18}
Sohlberg, K., Pennycook, S.J. \& Pantelides, S.T. Explanation of the
observed dearth of three-coordinated Al on {$\gamma$}-alumina surfaces.
{\it J. Am. Chem. Soc.} {\bf 121,} 10999--11001 (1999).

\bibitem{19}
Pennycook, S.J. \& Jesson, D.E. High-resolution incoherent imaging of
crystals. {\it Phys. Rev. Lett.} {\bf 64,} 938--941 (1990).

\bibitem{20}
Kirkland, E.J. {\it Advanced Computing in Electron Microscopy} (Plenum,
New York, 1998).

\bibitem{21}
Shannon, R.D. Revised effective ionic radii and systematic studies of
interatomic distances in halides and chalcogenides. {\it Acta Cryst. A}
{\bf 32,} 751--767 (1976).

\bibitem{22}
Burtin, P., Brunelle, J.P., Pijolat, M. \& Soustelle, M. Influence of
surface area and additives on the thermal stability of transition alumina
catalyst supports. I. Kinetic data. {\it Appl. Catal.} {\bf 34,} 225--238
(1987).

\bibitem{23}
Rashkeev, S.N. et al. Transition metal atoms on different alumina phases:
The role of subsurface sites on catalytic activity. {\it Phys. Rev. B}   
{\bf 67,} 115414 (2003).

\bibitem{24}
Krivanek, O.L., Dellby, N. \& Lupini, A.R. Towards sub-{\AA} electron
beams. {\it Ultramicroscopy} {\bf 78,} 1--11 (1999).

\bibitem{25}
Kresse, G. \& Furthm{\"u}ller, J. Efficiency of ab-initio total energy
calculations for metals and semiconductors using a plane-wave basis set.
{\it Comput. Mater. Sci.} {\bf 6,} 15--50 (1996).

\bibitem{26}
Perdew, J.P. Density-functional approximation for the correlation energy
of the inhomogeneous electron gas. {\it Phys. Rev. B} {\bf 33,} 8822--8824
(1986).

\bibitem{27}
Becke, A.D. Density-functional exchange-energy approximation with correct
asymptotic behavior. {\it Phys. Rev. A} {\bf 38,} 3098--3100 (1988).

\bibitem{28}
Vanderbilt, D. Soft self-consistent pseudopotentials in a generalized
eigenvalue formalism. {\it Phys. Rev. B} {\bf 41,} 7892--7895 (1990).

\bibitem{29}
Bl{\"o}chl, P.E. Projector augmented-wave method. {\it Phys. Rev. B} {\bf
50,} 17953--17979 (1994).

\bibitem{30}
Kresse, G. \& Joubert, D. From ultrasoft pseudopotentials to the projector
augmented-wave method. {\it Phys. Rev. B} {\bf 59,} 1758--1775 (1999).

\end{references}
\end{document}